\documentclass[aps,prl,twocolumn,letterpaper,superscriptaddress,showpacs]{revtex4-2}
\usepackage{amsmath, amsthm}
\usepackage{keyval}%
\usepackage{graphicx,latexsym}
\usepackage{dcolumn}
\usepackage{bm}
\usepackage{color}
\usepackage{url}

\usepackage{appendix}


\begin{document}

\title{Unusually stronger quantum fluctuation with larger spins: Novel phenomena revealed by emergent magnetism in pressurized high-temperature superconductor FeSe}


\author{Yuting Tan }
\affiliation{Department of Physics \& National High Magnetic Field
Laboratory, Florida State University, Tallahassee, FL 32306, USA}

\author{Tianyu Zhang}
\affiliation{Tsung-Dao Lee Institute \& School of Physics and Astronomy, Shanghai Jiao Tong University, Shanghai 200240, China}

\author{Tao Zou}
\affiliation{Department of Physics and Astronomy, Michigan State University, East Lansing, Michigan 48824-2320, U.S.A.}

\author{A. M. dos Santos}
\affiliation{Quantum Condensed Matter Division, Oak Ridge National Laboratory, Oak Ridge, Tennessee 37831, U.S.A.}

\author{Jin Hu}
\affiliation{Department of Physics and Engineering Physics, Tulane University, New Orleans, Louisiana 70118, U.S.A.}

\author{Dao-Xin Yao }
\affiliation{State Key Laboratory of Optoelectronic Materials and Technologies, School of Physics, Sun Yat-Sen University, Guangzhou 510275, China}

\author{Z. Q. Mao}
\affiliation{Department of Physics and Engineering Physics, Tulane University, New Orleans, Louisiana 70118, U.S.A.}

\author{Xianglin Ke}
\affiliation{Department of Physics and Astronomy, Michigan State University, East Lansing, Michigan 48824-2320, U.S.A.}

\author{Wei Ku }
\altaffiliation{corresponding email: weiku@sjtu.edu.cn}
\affiliation{Tsung-Dao Lee Institute \& School of Physics and Astronomy, Shanghai Jiao Tong University, Shanghai 200240, China}
\affiliation{Key Laboratory of Artificial Structures and Quantum Control (Ministry of Education), Shanghai 200240, China}

\date{\today}

\begin{abstract}
A counter-intuitive enhancement of quantum fluctuation with larger spins, together with a few novel physical phenomena, is discovered in studying the recently observed emergent magnetism in high-temperature superconductor FeSe under pressure.
Starting with experimental crystalline structure from our high-pressure X-ray refinement, we analyze theoretically the stability of the magnetically \textit{ordered} state with a realistic spin-fermion model.
We find surprisingly that in comparison with the magnetically ordered Fe-pnictides, the larger spins in FeSe suffer even stronger long-range quantum fluctuation that diminishes their ordering at ambient pressure.
This "fail-to-order" quantum spin liquid state then develops into an ordered state above 1GPa due to weakened fluctuation accompanying the reduction of anion height and carrier density.
The ordering further benefits from the ferro-orbital order and shows the observed enhancement around 1GPa.
We further clarify the controversial nature of magnetism and its interplay with nematicity in FeSe in the same unified picture for all Fe-based superconductors.
In addition, the versatile itinerant carriers produce interesting correlated metal behavior in a \textit{large} region of phase space. 
Our study establishes a generic exotic paradigm of stronger quantum fluctuation with larger spins that complements the standard knowledge of insulating magnetism.

\end{abstract}

\maketitle

Recent observation of Majorana zero mode~\cite{FeSe_zero_mode_dinghong} in single-layer FeSe thin film, a topological superconductor, has refueled the intense research interest in studying FeSe.
Particularly, the unusual high transition temperature (as high as 100K~\cite{FeSe_100K}) in this system and in the related compound K$_{0.8}$Fe$_{1.6}$Se$_2$ family~\cite{K0.8Fe2Se2} remains a challenging puzzle to our understanding.
While the exact microscopic mechanism of the high-$T_c$ superconductivity in these materials remains unclear, the proximity to long-range orders (in this case, antiferromagnetic order and nematic/ferro-orbital (FO) order~\cite{CCLee,Singh,Kruger}) implies that the strong correlations behind these orders are intimately related to the formation of high-$T_c$ superconductivity.
From this point of view, the FeSe families are particularly peculiar since FeSe and its isovalent families FeSe$_{1-x}$Te$_x$ and K$_{0.8}$Fe$_{1.6}$Se$_2$ present rich magnetic correlations, from the $C$-type~\cite{FeSe_Normal_state_JunZhao_ncomm, C_type_Nphysics}, $E$-type~\cite{E_type_Bao, E_type_Dai}, to the block~\cite{KFeSe_block_Bao} antiferromagnetic correlation with rather large fluctuating local moments~\cite{FeSeTeSet_Large_local_moment}, in great contrast to the rather consistent occurrence of a $C$-type antiferromagnetic correlation in nearly all Fe-pnictides~\cite{c_type_1111_nature_Dai, C_type_122_prl_Bao}.
Such a rich behavior is believed to originate from the strong interplay between itinerant carriers and the large local moments~\cite{Spin_fermion_Yin, Spin_fermion_Tan}.

\begin{figure}[t]
\begin{center}
\resizebox*{1.0\columnwidth}{!}{\includegraphics{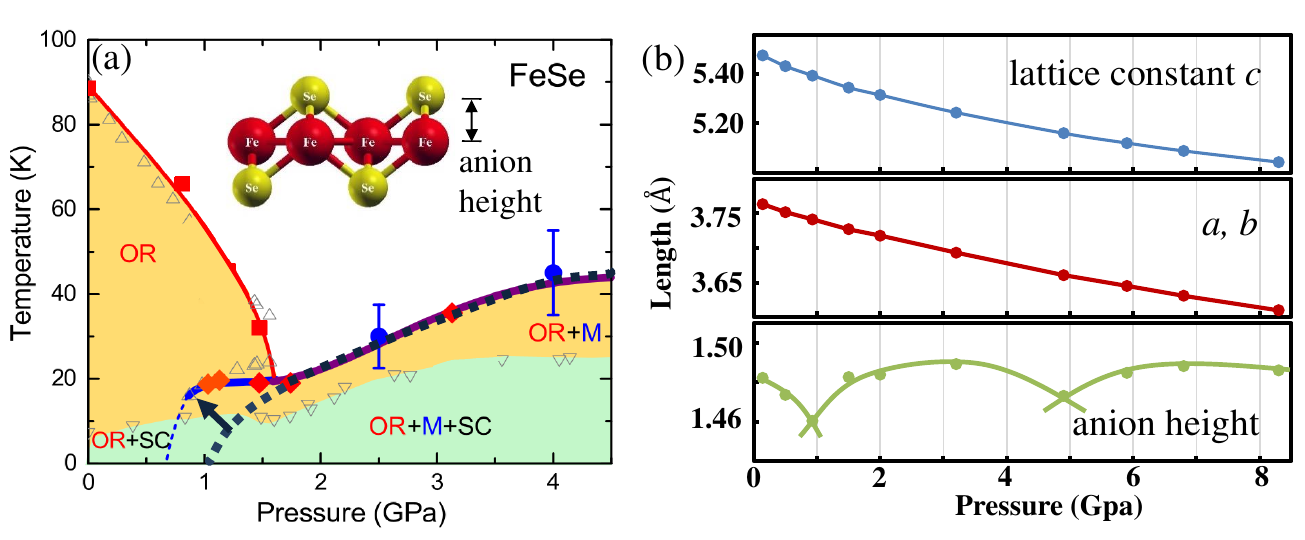}}
\end{center}
\caption{(a) Experimental phase diagram of FeSe against pressure~\cite{FeSe_part_phasediagram}, showing magnetic order above 1 GPa; (b) Experimental lattice structure from our X-ray refinement at various pressures(The error bar is smaller than the size of the symbol).  Notice a clear enhancement of magnetic order and a change of trend in the anion height (defined in the insect) inside the structural (ferro-orbital) ordered phase below 1.5GPa.}
\label{fig:fig1}
\end{figure}

Even more exotic is the lack of magnetic order in the pure FeSe, despite its large local moment almost twice the value of those in the magnetically ordered Fe-pnictides.
This contradicts directly the common lore of a weaker fluctuation with larger moments.
On the other hand, a very strong nematic/ferro-orbital order still remains in FeSe.
This ``departing'' of the magnetic and ferro-orbital degrees of freedom is in great contrast to the rather close proximity of these two orders in most Fe-pnictides.
It questions the validity of spin-nematicity being the leading physical effect in causing the nematic order~\cite{IIMazin, Nematic_JunZhao_nmaterial, Fernandas}.
This issue gains even more attention with the recent finding of pressure induced magnetic order in FeSe.
As shown in Fig.~\ref{fig:fig1}(a), the magnetic order emerges upon 1GPa of pressure and form a dome shape on its own~\cite{FeSe_1st_exp_pres, FeSe_part_phasediagram, FeSe_whole_phasediagram}.
While it is possible to mimic this unexpected behavior via perturbation theory~\cite{Kontani_Nematicity}, the two independent domes of the ferro-orbital order and the magnetic order strongly suggest their independent origins, and call for a physical picture for 1) Why is FeSe magnetically disordered at low pressure even though it has a much larger local moment than the pnictides, 2) How is FeSe different from Fe-pnictides that order magnetically, 3) How does magnetic order emerge in FeSe at higher pressure, and 4) How does ferro-orbital order interplay with the magnetic order?
Answering these questions would not only clarify the unique physics observed in FeSe, but also offer a generic unified physical picture for the electronic correlations in all Fe-based superconductors.

Here, we investigate these issues via a magnetically $ordered$-state stability analysis using a realistic spin-fermion model obtained from our experimental lattice structure refinement at various pressures.
Several surprising novel physical phenomena are discovered, resulting from the rich interplay between the large local moments and the itinerant carriers in FeSe.
First, at low pressure, the magnetically ordered state is destroyed by strong long-range quantum fluctuation originating from the itinerant carriers, despite the fact that FeSe has much larger spins than those magnetically ordered Fe-pnictides.
This novel behavior is related to a stronger polarization of itinerant carriers by larger spins, and is opposite to the rule of thumb that quantum fluctuation is only relevant to systems with small spins.
Second, under increased pressure, the fluctuation weakens systematically, accompanying the reduction of anion height and carrier density, and the magnetic order emerges from a ``fail-to-order'' quantum spin liquid state around 1GPa, where ferro-orbital order gives rise to the observed enhanced order.
Third, our study further clarifies how in essence the orbitals fluctuate differently from the spins, which naturally leads to two separate domes in the phase diagram.
Finally, the versatile itinerant carriers create a correlated metal state in a rather large region of the phase diagram, a feature of general interest to not only the Fe-based superconductors, but many strongly correlated systems as well.
Our study provides a new paradigm of large-spin fluctuation that complements the current lore of insulating magnetism.

To capture the realistic pressure dependence of the system, we first measure the structural parameters, lattice constants and atomic positions, of the material under various pressure. Room-temperature high-pressure synchrotron powder diffraction measurements were carried out at beamline 17-BM at the Advanced Photon Source using a monochromatic x-ray beam at a wavelength of $\lambda = 0.727750 \AA$ collimated down to $100 {\mu}m$ in diameter. Variable pressure diffraction data were collected in-situ using a Perkin Elmer amorphous-Si flat panel detector centered on the x-ray beam. The sample-to-detector distance was nominally set at $300 mm$, yielding an available $2\theta$ scattering angle of $27.5^{\circ}$, corresponding to access of Bragg reflections with d-spacing as low as $1.52 \AA$. The diffractometer geometrical parameters (such as precise sample-to-detector distance and tilt of the detector) were optimized with respect to a NIST a $LaB_6$ (660a) standard. The FeSe powdered sample was loaded in a membrane driven diamond anvil cell fitted with a pair of $800 {\mu}m$ culet diamonds. Gold was added to the sample as a pressure manometer and silicone oil was used as pressure-transmitting medium. The pressure values were determined by fitting the measured gold unit cell volume to a third-order Birch-Murnaghan equation of state using the parameters $V_0 = 67.850 {\AA}^3$, $K_0 = 167$GPa, and  $K'= 5.5$\cite{Dion},  where $V_0$ is the Au primitive cell volume at ambient conditions, $K_0$ is the bulk modulus, and $K'$ is its first pressure derivative. Rietveld refinement using Fullprof software package\cite{J.Rod}  of the x-ray powder diffraction were performed to determine the sample's crystallographic structure and lattice parameters under applied pressure.

Figure~\ref{fig:fig1}(b) gives the results of our structural refinement.
While the lattice parameters (top two panels) decrease smoothly under pressure, there is an obvious change of trends in the anion height (bottom panel) around 1GPa and 5GPa.
Unlike the lattice constant $c$ controlled by the Van der Waals interactions, the anion height is more sensitive to the local charge distribution.
It therefore correlates very well with the strength of ferro-orbital/structural correlation of the system.
From ambient pressure to 1GPa, as the ferro-orbital correlation grows weaker the anion height also becomes smaller.
Similarly between 1GPa and 5GPa, where magnetic correlation is strong~\cite{FeSe_two_domes}, the spin-nematicity and its induced ferro-orbital correlation also enhances, giving a larger anion height.
Note that the above (short-range) correlations must be driven by rather high energy physical mechanisms, since their influence on the anion height persists way above the transition temperature, where our measurement is performed.
Interestingly, beyond 5GPa, where superconducting correlation becomes stronger, the anion height again becomes larger, reflecting the enhanced short-range ferro-orbital/nematic correlation in the superconducting phase of Fe-based superconductors~\cite{Li111,LiFeAs_nematicity,Fe_nematicity,RbFe2As2_nematicity}.

We then use the experimental data as input to construct a simplest realistic spin-fermion Hamiltonian that incorporate both the local moment and the itinerant carriers~\cite{Spin_fermion_Yin, Spin_fermion_Lv, Spin_fermion_Weng, Spin_fermion_Dagotto, Spin_fermion_Tan}:
\begin{eqnarray}
\label{eq:eqn1}
\mathcal {H} &=& J_1\sum_{<i,i^\prime>}\vec{S}_i \cdot \vec{S}_{i^\prime} + J_2\sum_{<<i,i^\prime>>}\vec{S}_i \cdot \vec{S}_{i^\prime} \nonumber\\
 &-& J_H\sum_{i,m,s s^\prime}\vec{S}_i \cdot c_{i m s}^\dag \vec{\sigma}_{s s^\prime}c_{i m s^\prime}\\
 &-&\sum_{i i^\prime,n n^\prime, s}t_{i i^\prime}^{n n^\prime} c_{i n s}^\dag c_{i^\prime n^\prime s} + \sum_{i, n, s}(\epsilon\eta_n - \mu) c_{i n s}^\dag c_{i n s},\nonumber
\end{eqnarray}
in which the local moments $S_i$ at site $i$ couples antiferromagnetically to its first and second neighbors via $J_1$ and $J_2$, and couples ferromagnetically to the Fe-$d$ orbitals $m$ of the itinerant carriers via Hund's coupling $J_H$.
(Here $\vec{\sigma}_{ss^\prime}$ is a vector of the Pauli matrices.)
The itinerant carrier $c_{i n s}^\dag$ of Fe-$d$ and Se-$p$ orbitals $n$ and spin $s$ at site $i$ hops with pressure-dependent parameter $t_{ii^\prime}$.
Since Eq.~\ref{eq:eqn1} lacks interaction between fermions that drives the intrinsic orbital instability, we include the effect of FO order via an order parameter $\epsilon$ that splits the energy of $yz$ (upward) and $xz$ (downward) orbitals via $\eta_n$.
The chemical potential $\mu$ is evaluated for each set of magnetic and orbital order parameters to ensures equal number of electron and hole carriers, corresponding to the undoped parent compounds.

The pressure dependence of the hopping parameters $t_{ii^\prime}$ is obtained by applying the experimental structural data at each pressure to density functional theory (DFT) calculations of the normal-state, and then constructing symmetry-respecting Wannier functions~\cite{Ku2002} to extract the corresponding low-energy DFT hamiltonian in the tight-binding form.
It is important to note that with realistic $t_{ii^\prime}$ of FeSe, we found it necessary to incorporate the Se-$p$ orbital as well (an 8-band model) in order to reproduce the experimentally observed low-energy magnetic excitations.
While it is known that the quasi-particle excitation has rather stronger renormalization for some of the orbitals in FeSe~\cite{DFT_adjusted, ARPES_DFT_nematic_order_FeSe, ARPES_DFT_FeSe}, we find the use of bare $t$'s in our calculation produces quantitatively reasonable results, probably because the leading factor is the anisotropy of the Fermi surfaces, which is reserved under renormalization.
In the absence of experimental data, the coupling $J_1$ and $J_2$ between the local moments and their coupling to the itinerant carriers $J_H$ will be assumed to be weakly pressure dependent, in order to have a cleaner investigation of the physical effects without too many tuning parameters.
All the results discussed below thus correspond to bare couplings of the local moment $J_1=19meV$, $J_2=12meV$ and $J_H=0.8eV$ that produce the renormalized spin wave dispersion with the correct experimental band width~\cite{FeSe_Normal_state_JunZhao_ncomm} at low pressure.
This simplification can easily be improved with future experimental data, but should not alter the physical messages discussed below. 

Finally, we investigate the \textit{ordered} state stability via the linear spin wave theory by integrating out the fermion degrees of freedom~\cite{Spin_fermion_Lv,Spin_fermion_Tan} in the ordered state, up to the second order in $J_H$.
A discrete 500x500 momentum mesh and a thermal broadening of 10meV are used to ensure a good convergence.
The resulting renormalized spin wave Hamiltonian then gives the spin wave dispersion~\cite{Spin_fermion_Tan} that reflects the stability of the ordered state.

\begin{figure}[t]
\begin{center}
\resizebox*{1.0\columnwidth}{!}{\includegraphics{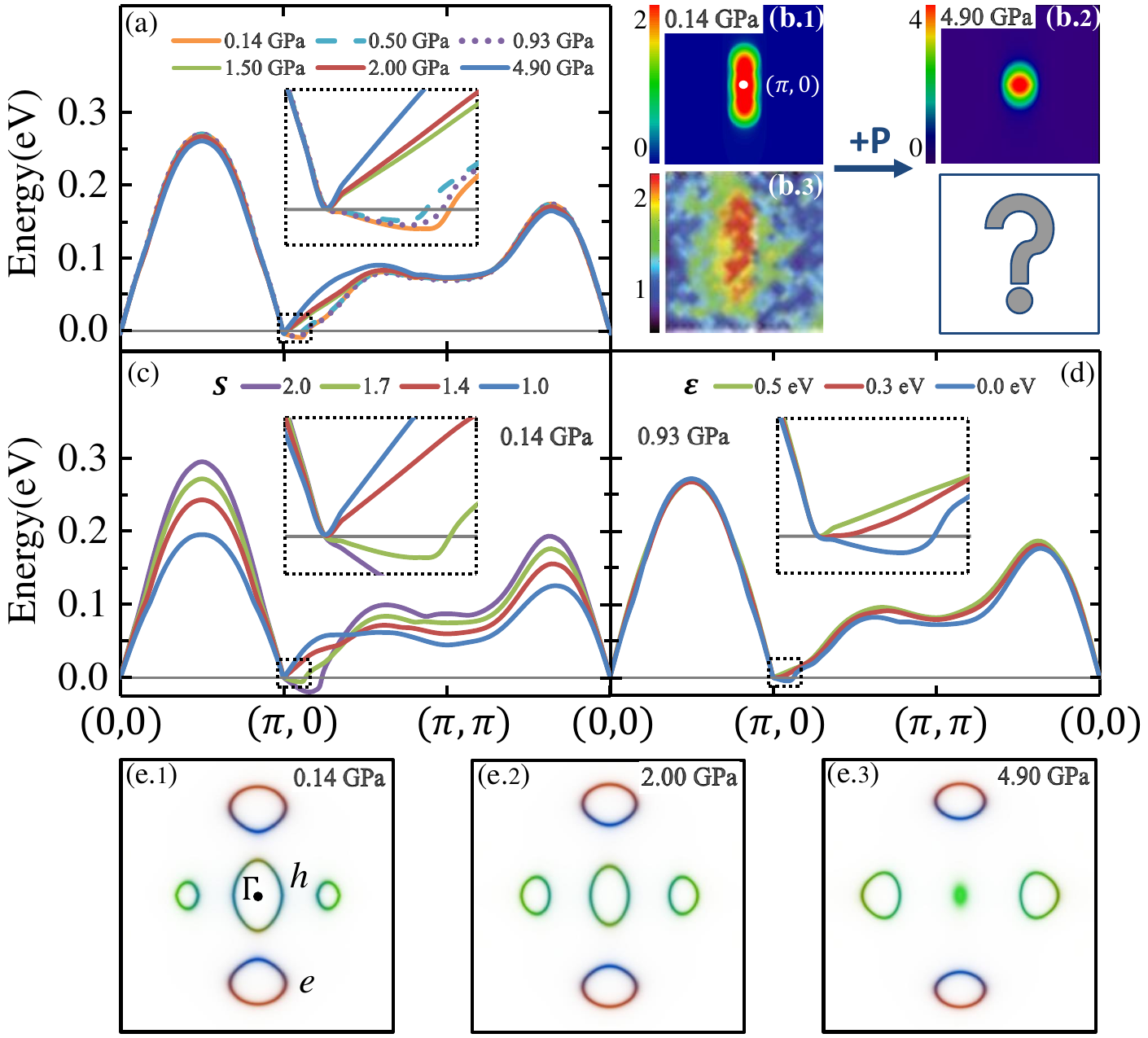}}
\end{center}
\caption{(Color online) Calculated spin wave dispersion of the magnetically ordered state along the high-symmetry paths: (a) pressure dependence showing stable ordered state above 1GPa, (c) local moment dependence at $P=0.14$GPa showing unstable ordered state beyond $S=1.5$, and (d) orbital order dependence at $P=0.93$GPa showing increased stability ordering by ferro-orbital order.
Here $S=1.7$ and $\epsilon=0$eV are used, unless specified explicitly.
(b) Momentum distribution of theoretical (upper panels) and experimental~\cite{C_type_Nphysics} (lower panels) spectra of low-energy magnetic excitation at 10$\pm 1$meV, for low-pressure fluctuating state (left panels), and high-pressure ordered state (right panels) in $\pm\pi/2$ momentum range around ($\pi$,0).
(e) Underlying ordered-state electron-, $e$, and hole-, $h$, Fermi surfaces entering our calculations, showing a reduction of carrier density (area of the pockets) at higher pressure (especially above 1.5GPa).
}
\label{fig:fig2}
\end{figure}

Figure~\ref{fig:fig2}(a) gives our spin wave dispersion of the \textit{ordered} state over the experimentally relevant pressure range 0.14GPa to 4.9GPa.
At low pressure, the obtained spin wave energy is imaginary (shown as negative in the standard convention) in the vicinity of ($\pi$,$q\sim 0$).
This indicates that the ($\pi$,0) ordered state is unstable against itinerant carrier-induced long-range quantum fluctuations~\cite{Spin_fermion_Tan} at the length scale longer than $2\pi/q$ along the $y$ direction, and thus the system would "fail to order", in agreement with the experimental observation.
This result provides a proper physical description of quantum fluctuation behind the recent density functional theory results~\cite{zyLu} that gives very similar total energy for systems with arbitrary mixing of ($\pi$,$2\pi/N$) with $N$=2 to 4.
By contrast, at higher pressure, the spin wave energy grows more positive;  The ($\pi$,0) order becomes stable beyond 1GPa, in excellent agreement with the experimental observation.

These results reveal the essential fluctuation-dominant nature of the magnetic order/disorder of FeSe.
The absence of magnetic order at low pressure is due to a strong itinerancy-induced long-range quantum fluctuation~\cite{Spin_fermion_Tan}, rather than the weakness of the normal-state instability typically derived from perturbation theories~\cite{Kotani,Scalapino_pertu}.
Particularly, besides the small momentum region near ($\pi$,0), our resulting spin wave dispersion is rather well-defined in nearly the entire momentum space, indicating that the short-range magnetic correlation remains very strong even in the absence of long-range order.
This is in good agreement with recent observation of magnetic excitation in the spin-disordered state~\cite{FeSe_Normal_state_JunZhao_ncomm}. Consistently at high pressure, given that the strong short-range correlation is already well established, the emergence of magnetic order simply follows the systematic reduction of the quantum fluctuation, rather than an enhancement of the normal-state instability.

The strong quantum fluctuation discussed here leaves a clear signature in the low-energy magnetic excitation that can be easily observed in inelastic neutron scattering experiments.
As shown in Fig.~\ref{fig:fig2}(b.1), at low pressure where fluctuation is so strong that long-range order is completely destroyed, the low-energy excitation shows a clear elongation along the $y$-direction near the magnetic Bragg peak at ($\pi$,0), reflecting the strong quantum fluctuation involving these momenta.
In fact, such elongation has been clearly observed in the previous measurements of the normal-state FeSe at 110K above the superconducting temperature~\cite{FeSe_Normal_state_JunZhao_ncomm}.
Of course, to really verify the quantum nature of the fluctuation, low-temperature observation is necessary.
Such data is available in FeTe$_{0.51}$Se$_{0.49}$~\cite{C_type_Nphysics}, which has very similar low-energy magnetic properties but negligible superconductivity.
Indeed, Fig.~\ref{fig:fig2}(b.3) demonstrates a clear elongation in FeTe$_{0.51}$Se$_{0.49}$ similar to our results.
Notice that the strong intensity near ($\pi$,0) is inconsistent with the proposal of quadrupolar order~\cite{QMSi}.
Also note that in our calculation this elongation greatly benefits from the already weak stiffness in the local moment system when $J_2$ is not much larger than $J_2/2$.
Therefore, we expect this elongation more robust than in the itinerant-only pictures~\cite{pure_itinerant_picture}.

On the other hand, at higher pressure, where fluctuation is weaker and long-range order emerges, the elongation is significantly reduced  [Fig.~\ref{fig:fig2}(b.2)].
Future experimental verification of this behavior would provide a strong support to our fluctuation dominant picture of the emerged magnetic order at high pressure.

Note that the inelastic neutron experiments~\cite{FeSe_Normal_state_JunZhao_ncomm} have actually ruled out the current proposals of competing orders~\cite{IIMazin,DHLee} as explanation for the lack of magnetic order at ambient pressure in FeSe.
Hardly any intensity below 10meV was found near ($\pi$,$\pi$) or ($\pi$,$\pi/2$) to generate meaningful quantum fluctuation capable of suppressing the magnetic order completely.
Recent NMR measurement~\cite{WeiqiangYu} also concluded no low-energy excitation beyond the vicinity of ($\pi$,0).
In fact generally speaking, production of unordered phase requires topological or geometrical ingredients beyond just the existence of competing orders of different momenta.
For example, unordered valence bond solid or spin liquid state can sometimes emerge between ordered states in geometrically frustrated magnetic systems, only when a \textit{continuous} supply of nearly degenerate ground states is available~\cite{FrustratedSpinSystems2013}, just as in our results along the ($\pi$,$q\sim0$)-line.
On the contrary, neutron experiments~\cite{FeSe_Normal_state_JunZhao_ncomm} found a 60meV strong dispersion from ($\pi$,0) to ($\pi$,$\pi$).
Therefore, had the energy at ($\pi$,$\pi$) been lowered to zero, the system would simply cross a quantum critical point and order in the new momentum, without an intermediate unordered phase.

Our analysis reveals two underlying factors for the reduction of quantum fluctuation under increased pressure.
First, the dramatic jump in the results across 1.5GPa [c.f. inset of Fig.~\ref{fig:fig2}(a)] coincides with that in the anion height in Fig.~\ref{fig:fig1} (which reflects the underlying nematic correlation as discussed above.)
In the tetrahedral coordination, the anion height is known to play an essential role in the electronic structure of Fe-based superconductors~\cite{Lin2011,anion_exp}, so it is not surprising that it also modifies the long-range fluctuation through tuning the collective screening of itinerant carriers.
Second, we observe a reduction of the itinerant carrier density in the ordered state, specially above 1.5GPa, consistent with the decreasing magnetic fluctuation.
(See for example the reduced electron pocket in the bottom panels of Fig.~\ref{fig:fig2}.)
Naturally, with smaller number of carriers, particularly corresponding to the hole pocket at $\Gamma$ point, their contribution to magnetic fluctuation would reduce.



So, why is FeSe different from the Fe-pnictides that have smaller local moment and yet still order magnetically?
The microscopic key factor turns out to be very surprising and against the conventional wisdom: its much larger local moment~\cite{FeSeTeSet_Large_local_moment, FeTe_localmoment}!
In conventional theory of insulating anti-ferromagnetic magnetism, the fluctuation originates from the spin-1 $S^+S^-$ flip processes between the neighboring spins, which becomes insignificant compared to the larger spins with stronger stiffness.
This consideration leads to the rule of thumb that larger spins fluctuate less and behave more classically.
Figure~\ref{fig:fig2}(c) shows a clear \textit{opposite} trend in our study.
As we tune the value of local moments from the smaller value comparable to the regular pnictides ($\sim 1$) to the larger value of FeSe ($\sim 1.7$), the resulting spin wave softens near ($\pi$,0), and the order becomes unstable beyond $S=1.4$.
Specifically, due to the coupling between the two degrees of freedom, a larger moment induces a stronger spin polarization of the itinerant carriers, which in turn fluctuate more the local moments at long range.
Since the itinerant carriers are more versatile against various ordering momenta, their enhanced fluctuation due to larger local moment also naturally leads to richer variation of ordering momenta in the FeSe$_{1-x}$Te$_x$ families~\cite{Spin_fermion_Yin}.
In essence, this mechanism provides a new paradigm opposite to the current lore of magnetism in spin-only systems.

Note that the sensitivity to the size of the local moment offers a second possible mechanism to give the observed dome shape~\cite{FeSe_whole_phasediagram} of the magnetic moment under pressure outside the scope of our study.
Starting at low pressure, where local moments are very large so does the fluctuation, applied pressure should systematically reduce the size of the local moments and thus reduce the fluctuation to eventually allow the long-range order to establish and to grow stronger.
Of course, this trend will eventually switch to a different one, where the local moments start to reduce significantly at high pressure, and the long-range order becomes weakened and eventually diminishes, forming the right side of the dome shape.
This mechanism might be relevant in the real material at higher pressure, and can be systematically investigated once the experimental quantification of the local moments become available at high pressure.


\begin{figure}[t]
\begin{center}
\resizebox*{1.0\columnwidth}{!}{\includegraphics{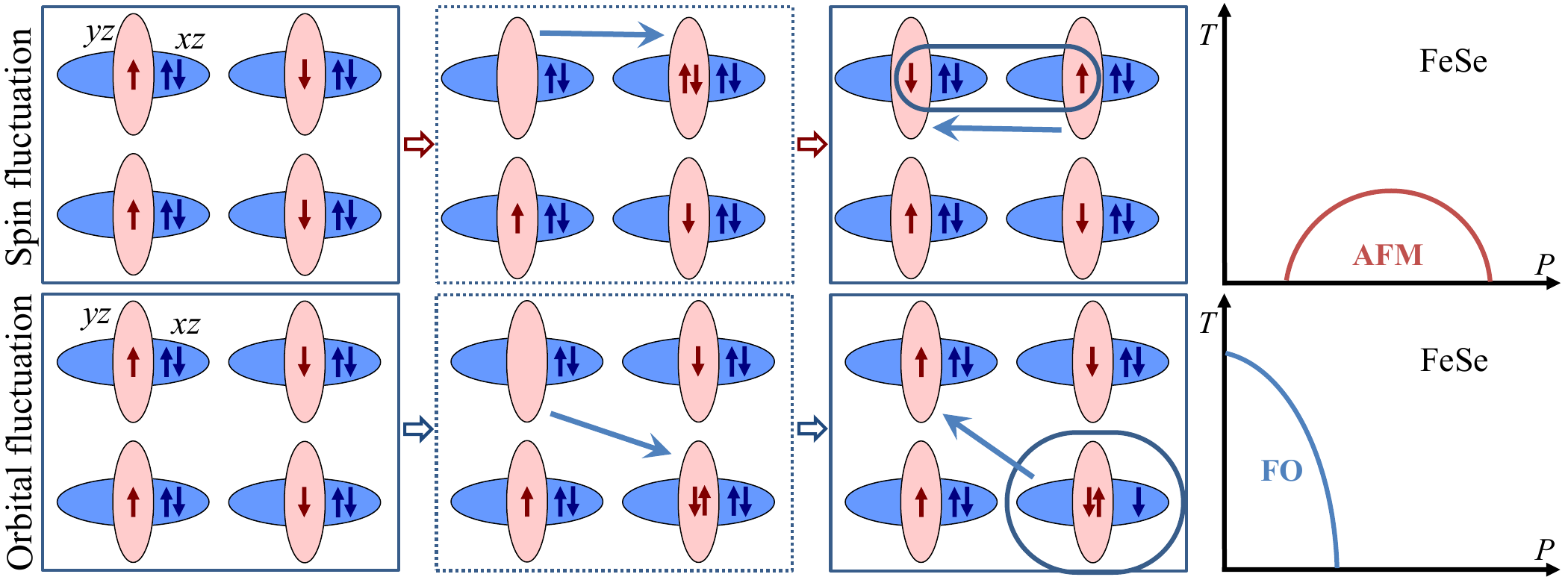}}
\end{center}
\caption{Illustration of distinct channels of spin- (upper panels) and orbital-fluctuations (lower panels) of an spin and orbital ordered state through virtual processes (middle panels).
Their different channels naturally lead to separate domes in the phase diagram (right most panels).}
\label{fig:fig3}
\end{figure}

We now clarify the heatedly debated issue of relationship of the structural/ferro-orbital/nematic order and the magnetic order.
Concerning the origin of the former, a popular viewpoint is the spin nematicity that completely ignores the orbital degree of freedom:
For the ($\pi$,0) magnetic order, Ising $Z_2$ symmetry must be broken prior to (or at the same time as) the rotational symmetry.
One thus expects the normal-state spin fluctuation to induce some degree of nematicity or even to stabilize its order~\cite{Fernandas}.
However, Fe-based superconductors have additional two-fold degenerate orbital freedom beyond the spin-only consideration, and thus a larger group structure is necessary.
In particular, the partially filled degenerate $xz$ and $yz$ orbitals are highly anisotropic in their coupling to the neighboring orbitals, and thus cannot be ``integrated out'' in any reasonable analysis of nematicity.
Unfortunately in most Fe-pnictides, the rather strong coupling between the orbital and spin correlations promotes a close proximity of their ordering, masking the dominant microscopic mechanisms.

The case of pressurized FeSe offers an unique example in which these two orders are well separated in the pressure range, as shown in Fig.~\ref{fig:fig1}(a).
Given that the magnetic order forms its own dome at high pressure, it is simply unreasonable to expect its associate spin nematicity~\cite{IIMazin,DHLee,Fernandas} to form a separate dome at a very different pressure region.
Figure~\ref{fig:fig2}(d) illustrate the impact of the ferro-orbital order on the magnetic fluctuation and ordering at 0.93GPa.
With enhanced ferro-orbital order, the spin wave energy hardens and the magnetic order becomes stabler and stronger.
This is in excellent agreement with the experimental observation of a clearly raised magnetic transition temperature $T_N$ around 1-1.7GPa in Fig.~\ref{fig:fig1}(a) (indicated by the arrow), in contrast to the smooth reduction from higher pressure.
If it were not with the extra help of ferro-orbital order, the magnetic order would have diminished around 1GPa.
In essence, the nematicity found in the magnetic order phase [the OR+M phase in Fig.~\ref{fig:fig1}(a)] is the true spin-fluctuation induced nematicity, which is obviously very weak, so weak that its ordering is indistinguishable from the full magnetic order even though theoretically it should form before the latter.
On the other hand, the much stronger nematicity at low pressure [the OR phase in Fig.~\ref{fig:fig1}(a)] is directly associated with the ferro-orbital order, and benefits little from spin nematicity in this case.

To further clarify the distinct nature of these two degrees of freedom and their ordering, Fig.~\ref{fig:fig3} illustrates a general process for their fluctuations in the atomic picture.
A spin and orbital well-ordered many-body state can couple to the spin fluctuated (upper panels) or orbital fluctuated (lower panels) states via virtual intermediate high-energy states of Hubbard-$U$ energy scale.
The stabilities of the long-range order of these two degrees of freedom are thus controlled by distinct fluctuation processes thermal- or quantum-wise, and are conceptually independent (even though they might couple to each other.)
Particularly, given the larger-moment-enhanced fluctuation in FeSe at low pressure, the unique separation of these two domes in the phase diagram of FeSe is no longer puzzling.
For completeness, the Appendix provides an illustration of the relationship between the AFM and FO domes in Fe-based superconductor families.

\begin{figure}[t]
\begin{center}
\resizebox*{1\columnwidth}{!}{\includegraphics{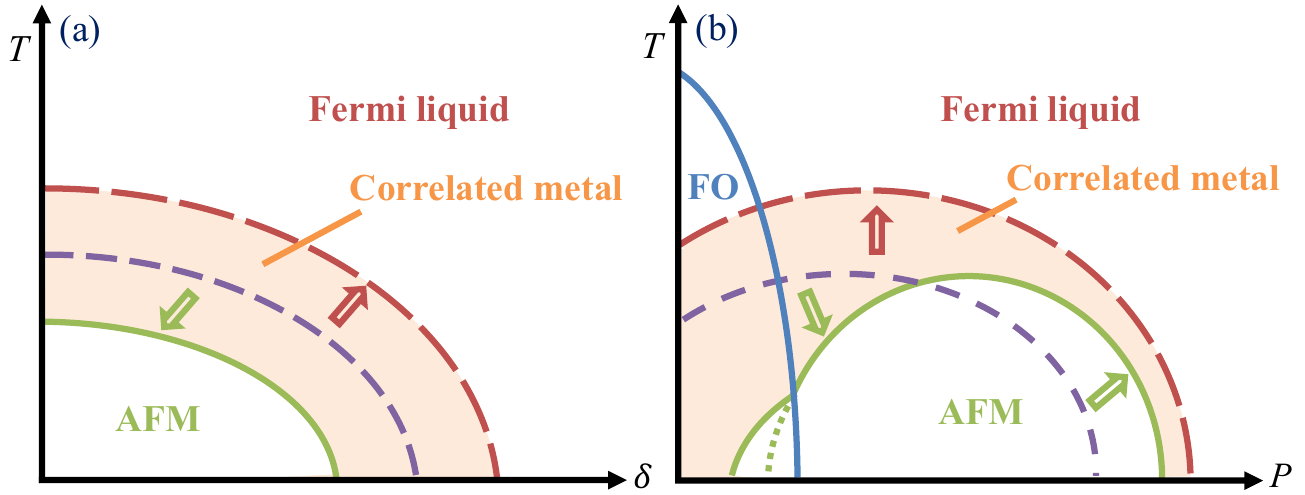}}
\end{center}
\caption{
Tendency to stretch the fluctuation region and emergence of correlated metal in Fe-pnictides (left) against doping $\delta$ and FeSe (right) against pressure $P$.
The versatile itinerant carriers tend to fluctuate the magnetic ordered state, moving its phase boundary from the spin-only system (dashed line) to that of the fermion-dressed system (solid line).
They also fluctuate the normal Fermi liquid state (long dashed line), creating a rather large region of a correlated metal with strong short-range correlation but without long-range order.
}
\label{fig:fig4}
\end{figure}

Finally, we notice an interesting emergence of a correlated metal behavior through the itinerant carrier induced fluctuation.
Start with only the local moments that order with a ($\pi$,0) momentum in part of the phase diagram, as marked by the dashed lines in Fig.~\ref{fig:fig4}(a) that illustrates the doping dependence of the pnictides.
Now adding coupling to the itinerant carriers with near ($\pi$,0)-nesting would naturally increase the instability of the normal state~\cite{itinerantcy_enhanced_stripe_Weng} and raise the temperature (red arrow) below which the normal Fermi liquid description is inadequate.
On the other hand, in the ordered state the same tendency toward destabilization by the itinerant carriers takes place as well and the real phase boundary is shifted to a lower temperature (green arrow).
This is due to the reconstruction of the itinerant Fermi surface under the broken symmetry, which gaps out all ($\pi$,0) nested bands and leaves only fluctuation in the vicinity of ($\pi$,0) that tends to destabilize the ordered state~\cite{Spin_fermion_Tan}.
As a result, a large part of the phase diagram is occupied by a correlated metal state that possesses strong short-range correlations but without long-range order.
This interesting consequence is generally applicable to systems with coupled local and itinerant degrees of freedom, for example also in the cuprates.

Fig.~\ref{fig:fig4}(b) shows a similar trend in FeSe at low pressure, but with an opposite trend at high pressure.
This is because the renormalization from the itinerant carriers no longer comes from the Fermi surface in this case, so the above consideration is altered.
Nevertheless, the strongly short-range correlated metal is still expected in a large region of the phase diagram, in which magnetic excitations should still be strong and can impact significantly physical properties, for example the electronic transport~\cite{JShen2009, ZXShen2011}.
Finally, this departing of the instability boundary of the normal Fermi liquid state from the actual phase boundary of ordering indicates a generic alarming inadequacy of the standard operation that detects the occurrence of low-energy magnetic, orbital, charge and pairing orders via the former (especially with low-order considerations).

In conclusion, several surprising novel physical phenomena are discovered, resulting from the rich interplay between the large local moments and the itinerant carriers in FeSe.
We analyze the magnetically $ordered$-state stability using a realistic spin-fermion model obtained from density functional calculation and our experimental lattice structure refinement at various pressures.
First, at low pressure, the magnetically ordered state is destroyed by strong long-range quantum fluctuation induced by the itinerant carriers, despite the fact that FeSe has much larger spins than those magnetically ordered Fe-pnictides.
This novel behavior is related to a stronger feedback screening of the $larger$ spins, and is opposite to the common lore that larger spins suffer less fluctuation.
Second, under pressure, the fluctuation weakens systematically accompanying the reduction of anion height and carrier density, and the magnetic order emerges from a "fail-to-order'' quantum spin liquid state around 1GPa, where ferro-orbital order gives rise to the observed enhanced order.
Third, our study further clarifies how in essence the orbitals fluctuate differently from the spins.
This naturally leads to two separate domes in the phase diagram, and resolves the highly debated issue of the underlying mechanism of nematicity in Fe-based superconductors.
Finally, the versatile itinerant carriers generate a correlated metal state in a rather large region of the phase diagram, a generic phenomenon of interest to a large class of strongly correlated materials.
Our study offers a new paradigm of large-spin quantum fluctuation that complements the current lore of insulating magnetism.

We thank V Dobrosavljevic, O Vafek, and Guangming Zhang for useful discussions.  YT and WK acknowledges support from National Natural Science Foundation of China \#11674220 and 11447601, and Ministry of Science and Technology \#2016YFA0300500 and 2016YFA0300501.
YT and DXY acknowledge support from NSFC-11574404, NSFG-2015A030313176, NSFC-Guangdong Joint Fund.
Portions of this research used resources at the Spallation Neutron Source, as well as the Advanced Photon Source, both user facilities operated by the Oak Ridge National Laboratory, and the Argonne National Laboratory respectively for the Office of Science of the U.S. Department of Energy (DOE).
Work at Michigan State University was supported by the National Science Foundation under Award No. DMR-1608752 and the start-up funds from Michigan State University.
Work at Tulane University was supported by the U.S. Department of Energy (DOE) under EPSCOR Grant No. DE-SC0012432 with additional support from the Louisiana Board of Regents (support for material synthesis).

\begin{appendices}
    \begin{center}
      {\bf APPENDIX}
    \end{center}
\subsection{Unusual elongation in hole dope Ba122}

\begin{figure}[t]
\begin{center}
\resizebox*{0.7\columnwidth}{!}{\includegraphics{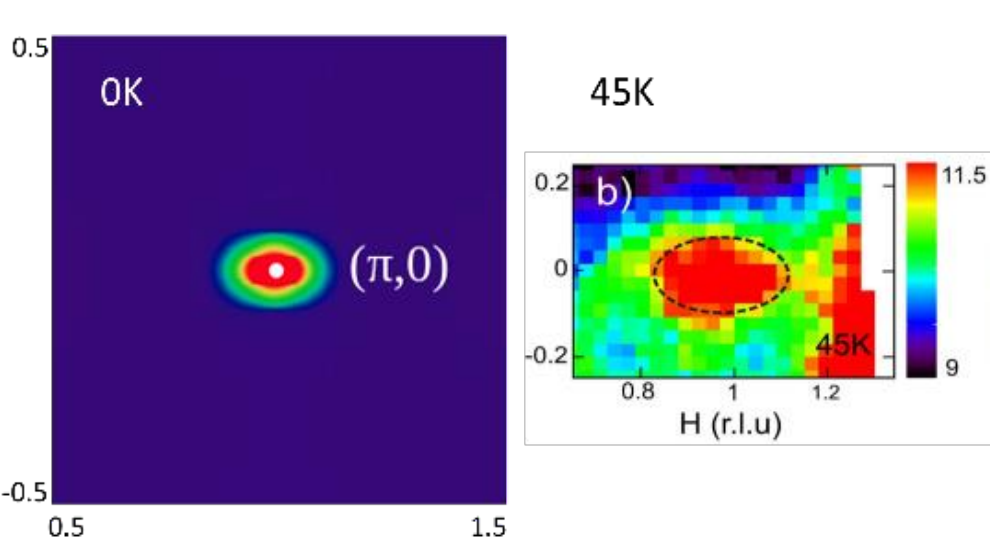}}
\end{center}
\caption{
(Color online) Momentum distribution of theoretical (left)  and experimental~\cite{Ba122holedoped} (right) spectra of low-energy magnetic excitation of Ba122 at 10$\pm 1$meV and 5$\sim$10meV respectively. 
}
\label{fig:fig5}
\end{figure}

Elongation along the ($\pi$,$q$)-line in Fe-based superconductors reflects the fluctuation, and is related to the soft stiffness of the local moments themselves.  It requires a significant change of the characteristics of the anisotropy of Fermi surface changes enough, as in Ba122 with enough hole doping, to switch the elongation to the ($\pi+q$,0)-line.  We perform the same calculation with input from undoped Ba122 with a shift of the chemical potential to simulate $26\%$ hole-doping.  The result is shown in the Fig.~\ref{fig:fig5}, and indeed now has an elongation along the ($\pi+q$,0)-line, similar to the experimental observation in the normal state~\cite{Ba122holedoped} . The higher temperature in the experiment will introduce additional thermal fluctuation, so the red area appears bigger.

\subsection{FO and AFM domes in different families}

The most relevant microscopic mechanisms of electronic nematicity are Ferro-orbital (FO) order, and the spin-nematic (SN) antiferromagnetic (AFM) phase (the Z2 symmetry broken ($\pi$,0) magnetic order).  Phase diagram of FeSe under pressure provides a solid experimental proof of the separation of the two domes associated with these two microscopic mechanisms of nematicity.  The same separation is also observed in Li111, in which a purely electronic (ferro-orbital FO driven) nematicity is observed~\cite{Li111}, while the magnetic ordered phase and its associate spin nematic (SN) dome is never found in ambient pressure.  This most likely can also be explained via strong long-range magnetic quantum fluctuation.

The AFM phase in FeSe is a good example that proves experimentally SN is generically very weak, so weak that its transition temperature is not distinguishable from that of the magnetic order (even though conceptually a strong SN should) through the entire pressure range.  That is, SN basically only coexists with the magnetic order and the nematicity outside the magnetic ordered phase is FO driven.

\begin{figure}[t]
\begin{center}
\resizebox*{1\columnwidth}{!}{\includegraphics{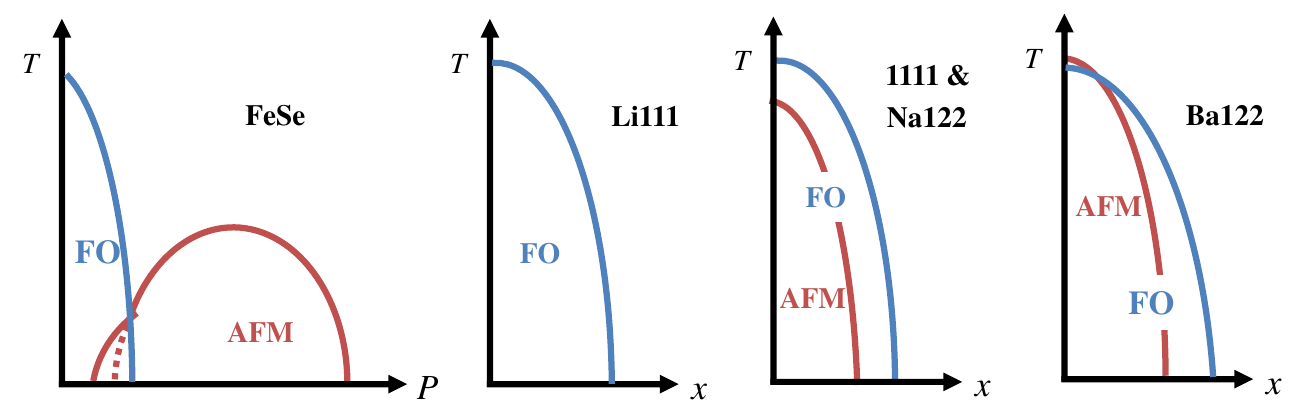}}
\end{center}
\caption{Illustration of the ferro-orbital order and anti-ferromagnetic order in Fe-based superconducting materials.
}
\label{fig:fig6}
\end{figure}

If one carries this conclusion to other Fe-based families, a very simple and universal picture would emerge as illustrated in Fig~\ref{fig:fig6}  In 1111~\cite{1111} and Na111~\cite{Na111}, the FO dome has higher transition temperature than the magnetic dome.  The same is true in Ba122 ~\cite{Na111} except at doping lower than $2\%$, where the FO thermal-fluctuates slightly stronger than the magnetic order.  In essence, since these two different mechanisms fluctuate with different microscopic processes (Fig.~\ref{fig:fig3}), they really should have individual domes in the phase diagram, even though these two mechanisms do couple to each other and slightly enhance each other~\cite{CCLee}.  

\end{appendices}

\bibliography{FeSe_v2}

\end{document}